\documentclass[aps,prl,reprint,superscriptaddress]{revtex4-2}

\usepackage{amsmath,amssymb,amsfonts,bm}
\usepackage{graphicx}
\usepackage{color}
\usepackage{subcaption}
\usepackage{xcolor}

\begin{document}

\title{Impurity-induced geometric correlations and fractional quantization in quantum Hall systems}

\author{M. A. Hidalgo}
\affiliation{Departamento de Física y Matemáticas, Universidad de Alcalá, Alcalá de Henares (Madrid), Spain}

\date{\today}

\begin{abstract}
We propose a geometric mechanism for fractional quantum Hall states based on impurity-induced correlations within a Landau level. A correlated distribution of ionized impurities partially modifies the Landau-level degeneracy through coherent coupling between cyclotron orbits, generating fractional energy sublevels. The odd-denominator hierarchy emerges naturally from the intrinsic guiding-center quantization and the correlated cyclotron motion. The resulting spectrum reproduces the principal experimentally observed fractional sequences and predicts a strong dependence of fractional-state stability on impurity geometry and layer separation. The absence of an incompressible Hall plateau at filling factor 1/2 follows from cancellation of the geometric correlations responsible for odd-denominator states. These results suggest that impurity-induced geometry may constitute an additional organizing principle in quantum Hall systems.
\end{abstract}

\maketitle

\section{Introduction}
The quantum Hall effect constitutes one of the clearest manifestations of quantum coherence in low-dimensional electronic systems. In a two-dimensional electron gas subjected to a strong perpendicular magnetic field, electronic motion becomes quantized into highly degenerate Landau levels, giving rise to the integer quantum Hall effect (IQHE) through Landau quantization and disorder-induced localization \cite{laughlin1981quantized}. The subsequent discovery of the fractional quantum Hall effect (FQHE) revealed incompressible electronic states at fractional filling factors \cite{laughlin1983anomalous}. The standard theoretical description is based on strongly correlated many-body states and emergent quasiparticles carrying fractional charge \cite{laughlin1983anomalous,jain1992microscopic}.

Despite the success of interaction-based approaches, several experimental features suggest that geometric aspects of Landau-level structure may also play an important role in fractional-state formation. In particular, the continuity between integer and fractional magnetotransport regimes and the strong dependence of fractional-state stability on sample quality and heterostructure geometry indicate that spatial correlations within a Landau level may contribute significantly to the observed hierarchy of fractional states.

In this work, we propose a geometric framework in which impurity-induced spatial correlations partially lift the Landau-level degeneracy. We consider a two-dimensional electron gas coupled to a correlated distribution of ionized impurities located at a finite distance from the electronic plane. The resulting impurity-induced orbital coherence generates effective fractional energy sublevels within a Landau level. In the present model, the odd-denominator hierarchy emerges naturally from guiding-center quantization and correlated cyclotron motion. The model reproduces the experimentally observed fractional sequences and predicts a direct dependence of fractional-state stability on impurity geometry and layer separation.

\section{LANDAU STATES AND IMPURITY-INDUCED GEOMETRIC CORRELATIONS}

We consider a two-dimensional electron system subjected to a perpendicular magnetic field $\mathbf{B}=B \hat{z}$. Within the effective-mass approximation and using the symmetric gauge, $\mathbf{A}=B (-y/2,x/2,0)$, the single-particle Hamiltonian is
\begin{equation}
H_0 = \frac{1}{2m^*} \left( \mathbf{p} + e\mathbf{A} \right)^2
\label{eq:H0}
\end{equation}
where $m^{*}$ is the effective electronic mass. The corresponding Landau eigenstates are
\begin{equation}
\psi_n^m(\mathbf{r},\theta) = C_{n,m} \, r^{|m|} e^{-r^2/4\ell_B^2} L_n^{|m|}\left(\frac{r^2}{2l_B^2}\right) e^{im\theta}
\label{eq:wave}
\end{equation}
with energies
\begin{equation}
E_n =\hbar\omega_c \left(n+\frac{1}{2}\right)
\end{equation}
where $\omega_c = eB/m^*$ and
\begin{equation}
l_B=\sqrt{\frac{\hbar}{eB}}
\end{equation}
is the magnetic length. 
In the symmetric gauge, the electron coordinates can be decomposed into cyclotron and guiding-center contributions. For a fixed Landau level, within the present phenomenological framework, the guiding-center radius is associated with an effective geometric quantization scale of the form \cite{cohen2019quantum1,haldane2011geometrical}
\begin{equation}
\langle r_{GC}^2 \rangle = q l_B^2
\label{eq:larmor}
\end{equation}
where $q$ is taken as an odd integer characterizing the correlated fractional states.
To investigate the effect of spatial correlations, we consider a correlated distribution of ionized impurities located at a finite distance $\Delta$ from the two-dimensional electron gas. The full Hamiltonian is
\begin{equation}
H = H_0 + U^{(i)}(\mathbf{r})
\label{eq:fullH}
\end{equation}
with impurity potential
\begin{equation}
    U^{(i)}(\mathbf{r})=-e^2\sum_j [|\mathbf{r}-\mathbf{R}_j|^2+\Delta^2]^{-1/2}
\end{equation}
In the high-field regime, where the magnetic length becomes comparable to the impurity-correlation scale, the continuous rotational symmetry of cyclotron motion  evolves into an effective correlated orbital structure. We therefore introduce a characteristic correlation length
\begin{equation}
|\boldsymbol{\xi}| = \eta d_i
\label{eq:xi}
\end{equation}
where $d_i$ is the dominant impurity spacing and $\eta$ is a dimensionless correlation parameter. The correlated orbital displacement is assumed to occur along an effective local correlation direction, $\boldsymbol{\xi} = \eta d_i \,\hat{\mathbf e}_{\xi}$, and the electronic states are described by correlated combinations of displaced Landau orbitals.
\begin{equation}
\Phi_{\xi}(k_q,\mathbf{r})=\frac{1}{\sqrt{N}}
\sum_{\alpha}
e^{ik_q\alpha\xi}
\psi_n^m(\mathbf{r}-\alpha\boldsymbol{\xi}),
\label{eq:Phi}
\end{equation}
which provide an effective phenomenological representation of impurity-induced orbital coherence within a correlation-modified Landau level. A schematic representation of these impurity-induced orbital correlations considered in the present model is shown in Fig. 1.
Since the guiding-center radius scales as \(\sqrt{q}\,l_B\), the characteristic correlation wave vector is
\begin{equation}
k_q=\frac{2\pi}{\sqrt{q}\,l_B},
\label{eq:kq}
\end{equation}
As shown below, impurity-induced coupling between correlated cyclotron orbits modifies the Landau-level degeneracy into a hierarchy of fractional energy states associated with the experimentally observed odd-denominator filling factors.

\section{EFFECTIVE FRACTIONAL ENERGY STRUCTURE}

The correlated basis introduced above modifies the Landau-level degeneracy through coherent coupling between displaced cyclotron states. Within the nearest-neighbor approximation, the dominant matrix elements are assumed to arise from orbitals separated by the correlation length $\xi=\eta d_i$,
\begin{equation}
\left\langle
\psi_n^m(\mathbf{r}+\alpha\boldsymbol{\xi})
\middle|H\middle|\psi_n^m(\mathbf{r}+[\alpha\pm1]\boldsymbol{\xi})
\right\rangle=\pm\frac{\gamma}{2}
\label{eq:coupling}
\end{equation}
where $\gamma$ is an effective impurity-induced correlation energy. Using the correlated states defined in Eq.~(\ref{eq:Phi}), the emergent spectrum becomes
\begin{equation}
E(k_q)=E_n\mp\gamma\cos(k_q\xi)
\label{eq:band}
\end{equation}
Retaining the leading correction to the cosine term yields
\begin{equation}
E(k_q)\simeq E_n \mp \gamma \pm \frac{\gamma}{2} (k_q\xi)^2.
\label{eq:expansion}
\end{equation}
Defining the shifted Landau energy
\begin{equation}
\tilde{E}_n = E_n\mp\gamma,
\end{equation}
and using Eq.~(\ref{eq:kq}) the impurity-induced contribution becomes
\begin{equation}
\Delta E_q^\eta=\frac{\gamma}{2} \left( \frac{2\pi\eta d_i}{\sqrt{q}l_B} \right)^2
\label{eq:deltaE}
\end{equation}
Now, introducing the correlation parameter
\begin{equation}
\eta=\sqrt{p}
\label{eq:p}
\end{equation}
where $p$ characterizes the dominant orbital correlation, the fractional energy spectrum can be written as
\begin{equation}
E=\left(2n+1\pm\frac{p}{q}\right)E_0\label{eq:fractionalE}
\end{equation}
with
\begin{equation}
E_0=\frac{\hbar\omega_c}{2}.
\end{equation}
Eq.~(\ref{eq:fractionalE}) constitutes the central result of the present work. The fractional hierarchy emerges from the interplay between guiding-center quantization and impurity-induced orbital correlations. The dominant experimentally observed odd-denominator states correspond to the shortest correlation lengths, namely  $\eta=1$ and \(\eta=\sqrt{3}\).
Fractional states appear as emergent sublevels of the original Landau spectrum. The absence of a robust incompressible state at filling factor 1/2 follows naturally from cancellation of the geometric correlations responsible for odd-denominator states ($p$=0).

\section{MAGNETOTRANSPORT AND COMPARISON WITH EXPERIMENT}

The impurity-induced splitting of the Landau spectrum modifies the density of states and generates additional minima in the longitudinal resistivity together with Hall plateaux at fractional filling factors. The spectrum is therefore obtained through the substitution
\begin{equation}
E\rightarrow\left(2n+1\pm\frac{p}{q}\right)E_0,
\label{eq:fractionalDOS}
\end{equation}
Shubnikov–de Haas minima occur when the Fermi level crosses the fractional sublevels, leading to the magnetic-field sequence
\begin{equation}
B_{n,p,q}\simeq\frac{2m^*E_F}{\hbar e
\left(2n+1\pm\frac{p}{q}\right)},
\label{eq:Bpq}
\end{equation}
where
\begin{equation}
E_F=\frac{2\pi\hbar^2 n_e}{m^*}
\end{equation}
is the Fermi energy of the two-dimensional electron gas with electron density $n_e$. Using experimentally realistic GaAs/AlGaAs densities, $n_e \simeq 3 \times 10^{15}\,\mathrm{m}^{-2}$ \cite{willett1987observation}, the resulting spectrum reproduces the principal experimentally observed odd-denominator sequences, including the dominant 1/3, 2/3, 2/5, and 3/5 states \cite{willett1987observation,clark1986odd,willett1988termination,du1993experimental,choi2008activation,shabani2010fractional}. The fractional sublevel structure associated with the 1/3 and 1/5 families is illustrated in Figs. 2 and 3, respectively.
The characteristic fractional energy scale is approximately
\begin{equation}
\Delta E_{p/q}\sim\hbar\omega_c\frac{p}{q}
\label{eq:gap}
\end{equation}
leading to the observability condition
\begin{equation}
\frac{p}{q}\gtrsim\frac{\Gamma}{\hbar\omega_c}
\label{eq:criterion}
\end{equation}
where $\Gamma$ denotes the disorder broadening. Eq.~(\ref{eq:criterion}) naturally explains the predominance of low-order odd-denominator fractions and the suppression of higher-order states as disorder increases.

\section{DISCUSSION}
The present results suggest that impurity-induced geometric correlations can partially reorganize the Landau-level degeneracy and contribute to the emergence of fractional quantum Hall states. In the present picture, the stability of fractional states is directly linked to impurity correlations and heterostructure geometry, particularly to the separation $\Delta$ between the impurity layer and the two-dimensional electron gas.

The proposed mechanism is phenomenological and does not explicitly incorporate electron–electron interactions or topological excitations such as fractional charge and braiding statistics. Consequently, the present approach is not intended to replace interaction-based descriptions of the fractional quantum Hall effect, but rather to identify a complementary geometric contribution associated with guiding-center coherence and impurity-induced orbital correlations.

The model also suggests the possibility of engineering aspects of fractional quantum Hall phenomenology through controlled impurity distributions and heterostructure design. More generally, these results indicate that geometry, disorder, and guiding-center coherence may play a broader role in the organization of highly degenerate Landau-level states.

\section{CONCLUSION}

We have presented a geometric framework for fractional quantum Hall states based on impurity-induced correlations within a Landau level. A correlated impurity environment generates coherent coupling between cyclotron orbits and provides fractional sublevels associated with odd-denominator filling factors.

The resulting spectrum reproduces the principal experimentally observed fractional sequences and predicts a direct dependence of fractional-state stability on impurity geometry, disorder correlations, and layer separation. The absence of an incompressible Hall plateau at filling factor 1/2 follows from cancellation of the geometric correlations responsible for odd-denominator states.

These results suggest that impurity-induced geometry and guiding-center coherence may constitute an additional organizing principle underlying quantum Hall phenomenology.

\begin{figure}[t]
\includegraphics[width=0.9\columnwidth]{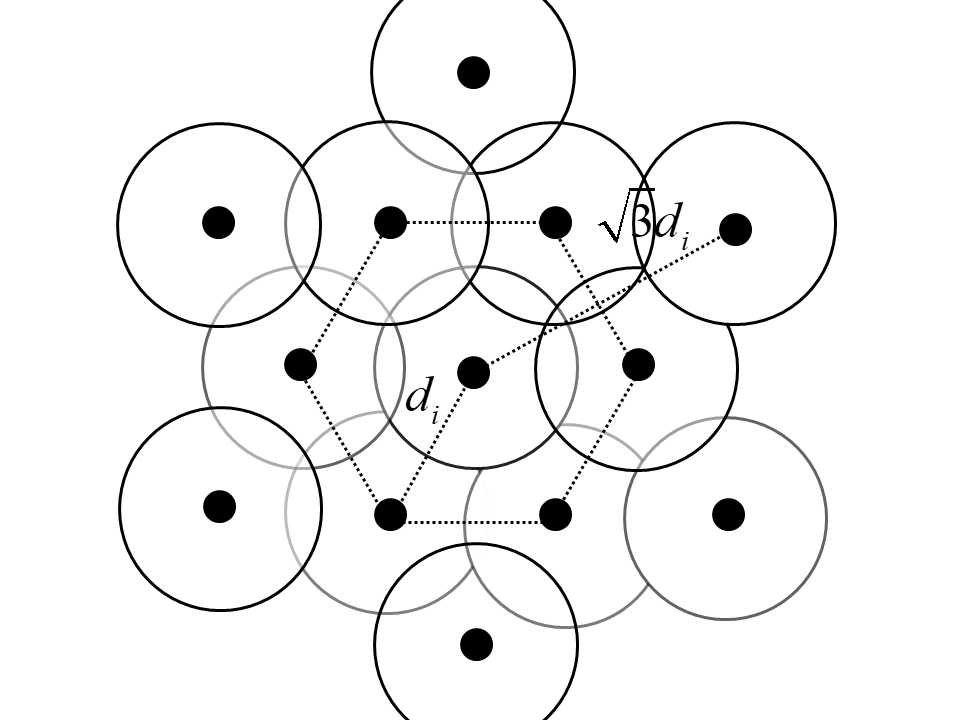}
\caption{Schematic representation of impurity-induced correlated cyclotron orbits in a 2DES.}
\label{fig:fig1}
\end{figure}

\begin{figure}[t]
\includegraphics[width=0.9\columnwidth]{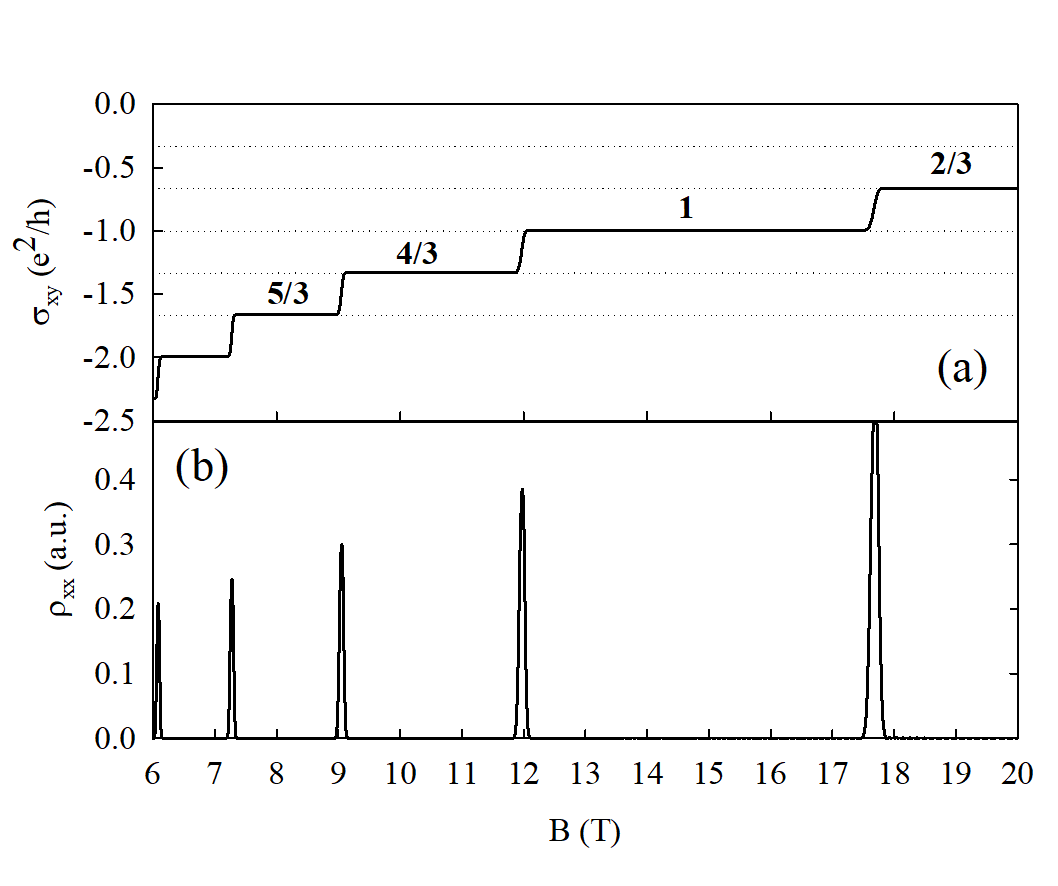}
\caption{(a) Hall magnetoconductivity $\sigma_{xy}$ showing the fractional plateaux associated with the 1/3 family. (b) Longitudinal magnetoresistivity $\rho_{xx}$ (Shubnikov–de Haas response) exhibiting the corresponding fractional minima predicted from the fractional spectrum obtained from Eq.~(\ref{eq:fractionalE}).}
\label{fig:fig2}
\end{figure}

\begin{figure}[t]
\includegraphics[width=0.9\columnwidth]{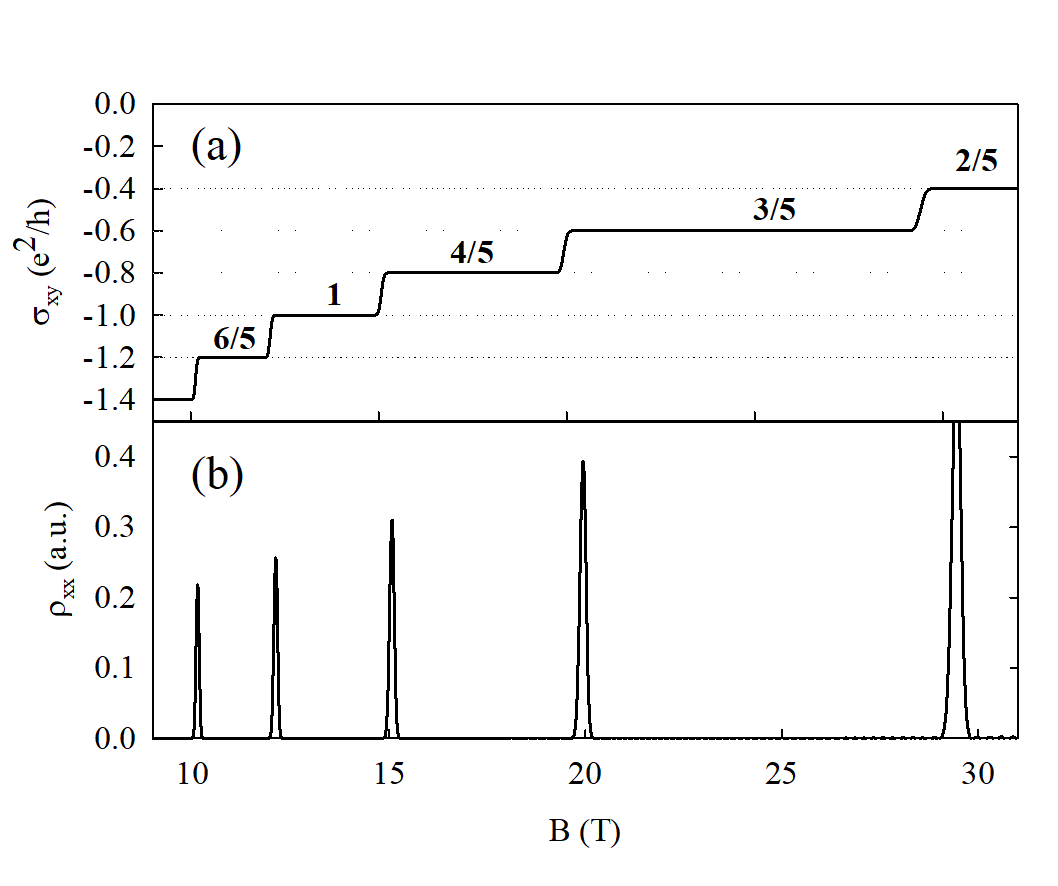}
\caption{(a) Hall magnetoconductivity $\sigma_{xy}$ showing the fractional plateaux associated with the 1/5 family. (b) Longitudinal magnetoresistivity $\rho_{xx}$ (Shubnikov–de Haas response) exhibiting the corresponding fractional minima predicted from the fractional spectrum obtained from Eq.~(\ref{eq:fractionalE}).}
\label{fig:fig3}
\end{figure}

\bibliographystyle{apsrev4-2}
\bibliography{references}

\end{document}